\documentclass{emulateapj}

\newcommand{\etal}{et~al.}
\newcommand{\eg}{e.g., }
\newcommand{\ie}{i.e., }
\newcommand{\Msun}{M_{\odot}}

\newcommand{\Nifs}{$^{56}$Ni}

\newcommand{\Ed}{\dot{E}_{\rm dep}}
\newcommand{\Edep}{\dot{E}_{\rm dep, 51}}

\newcommand{\Mni}{M{\rm (^{56}Ni)}}

\def\gsim{\mathrel{\rlap{\lower 4pt \hbox{\hskip 1pt $\sim$}}\raise 1pt
\hbox {$>$}}}
\def\lsim{\mathrel{\rlap{\lower 4pt \hbox{\hskip 1pt $\sim$}}\raise 1pt
\hbox {$<$}}}

\begin{document}

\title{The Connection between Gamma-Ray Bursts and Extremely Metal-Poor Stars:
Black Hole-forming Supernovae with Relativistic Jets}

\author{
 Nozomu~Tominaga\altaffilmark{1},
 Keiichi~Maeda\altaffilmark{2}, 
 Hideyuki~Umeda\altaffilmark{1},
 Ken'ichi~Nomoto\altaffilmark{1,3}, 
 Masaomi~Tanaka\altaffilmark{1},
 Nobuyuki~Iwamoto\altaffilmark{4},
 Tomoharu~Suzuki\altaffilmark{1}, and
 Paolo~A.~Mazzali\altaffilmark{1,5,6}
 }

\altaffiltext{1}{Department of Astronomy, School of Science,
University of Tokyo, Bunkyo-ku, Tokyo, Japan;
tominaga@astron.s.u-tokyo.ac.jp}
\altaffiltext{2}{Department of Earth Science and Astronomy, 
Graduate School of Arts and Science, University of Tokyo, Tokyo, Japan}
\altaffiltext{3}{Research Center for the Early Universe, School of
Science, University of Tokyo, Bunkyo-ku, Tokyo, Japan}
\altaffiltext{4}{Nuclear Data Center, Nuclear Science and Engineering
Directorate, Japan Atomic Energy Agency, Tokai, Ibaraki, Japan}
\altaffiltext{5}{Max-Planck Institut f\"ur Astrophysik,
Karl-Schwarzschild Strasse 1, D-85748 Garching, Germany}
\altaffiltext{6}{Istituto Nazionale di Astrofisica, Osservatorio
Astronomico di Trieste, Trieste, Italy}

\begin{abstract}
 Long-duration gamma-ray bursts (GRBs) are thought to be connected
 to luminous and energetic supernovae (SNe), called hypernovae (HNe),
 resulting from the black-hole (BH) forming collapse of massive
 stars. For recent nearby GRBs~060505 and 060614, however, the expected
 SNe have not been detected. The upper limits to the SN brightness are
 about 100 times fainter than GRB-associated HNe (GRB-HNe),
 corresponding to the upper limits to the ejected \Nifs\ masses of
 $\Mni\sim 10^{-3}\Msun$. SNe with a small amount of \Nifs\ ejection are
 observed as faint Type II SNe. HNe and faint SNe are thought to be
 responsible for the formation of extremely metal-poor (EMP) stars. In
 this Letter, a relativistic jet-induced BH forming explosion of the 40
 $\Msun$ star is investigated and hydrodynamic and nucleosynthetic
 models are presented. These models can explain both GRB-HNe and GRBs
 without bright SNe in a unified manner. Their connection to EMP
 stars is also discussed. We suggest that GRBs without bright SNe are
 likely to synthesize $\Mni\sim 10^{-4}$ to $10^{-3}\Msun$ or $\sim 10^{-6}\Msun$.
\end{abstract}


\section{INTRODUCTION}
\label{sec:introduction}

Long-duration gamma-ray bursts (GRBs) at sufficiently close distances
($z<0.2$) have been found to be accompanied by luminous and energetic
Type Ic supernovae (SNe Ic) called hypernovae (HNe; GRB~980425/SN~1998bw: \citealt{gal98}; GRB~030329/SN~2003dh:
\citealt{sta03}, \citealt{hjo03}; GRB~031203/SN~2003lw: \citealt{mal04}). 
These GRB-associated HNe (GRB-HNe) are suggested to be the outcome
of very energetic black hole (BH) forming explosions of massive stars
(\eg \citealt{iwa98}). The ejected \Nifs\ mass is estimated to be
$\Mni\sim0.3-0.7\Msun$ (\eg \citealt{maz06a}) which is $4-10$ times
larger than typical SNe Ic [$\Mni\sim 0.07\Msun$; \citealt{nom06}]. 

For recently discovered nearby long-duration GRB~060505 ($z=0.089$,
\citealt{fyn06}) and GRB~060614 ($z=0.125$,
\citealt{gal06,fyn06,del06,geh06}), no SN was detected.\footnote[7]{There is
a suspicion on whether the GRBs are classical
long-duration GRBs (GRB~060505: \citealt{ama06}, and GRB~060614:
\citealt{geh06}).} Upper limits to brightness of the possible SNe
are about 100 times fainter than SN~1998bw [\ie $\Mni\lsim 10^{-3}\Msun$]. 

A small amount of \Nifs\ ejection has been indicated in the faintness of
several Type II SNe (SNe II, \eg SN~1994W, \citealt{sol98}; and
SN~1997D, \citealt{tur98}). The estimated explosion energies, $E$, of
these faint SNe~II are very small ($E\lsim10^{51}$ergs,
\citealt{tur98}). These properties are well-reproduced by the {\sl
spherical} explosion models that undergo significant fallback if $E$ is
sufficiently small \citep{woo95,iwa05}. Thus these faint SN explosions
with low $E$ seem to be superficially irreconcilable to the formation of
energetic GRBs \citep{gal06}. However, Nomoto et al. (2006a;\footnote[8]{The ppt
file is available from ``program'' at
http://www.me- rate.mi.astro.it/docM/OAB/Research/SWIFT/sanservolo2006/}
see also \citealt{nom04}) has predicted the existence of ``dark hypernovae'' (i.e.,
long GRBs with no SNe) based on the argument in the next paragraph.  In
this Letter, we present actual hydrodynamical models
in which a high-energy narrow jet produces a GRB {\sl and} a faint/dark SN
with little \Nifs\ ejection.

An indication that some faint SNe produce high-energy jets is
seen in the abundance patterns of the extremely metal-poor
(EMP) stars with [Fe/H] $<-3.5$.\footnote[9]{Here [A/B] 
$\equiv\log_{10}(N_{\rm A}/N_{\rm B})-\log_{10}(N_{\rm A}/N_{\rm B})_\odot$,
where the subscript $\odot$ refers to the solar value and $N_{\rm A}$
and $N_{\rm B}$ are the abundances of elements A and B, respectively.}
It has been suggested that the abundance patterns of these EMP stars
show the nucleosynthesis yields of a single core-collapse SN (\eg
\citealt{bee05}). In particular, the C-enhanced type of the EMP stars have
been well explained by the faint SNe \citep{ume05,iwa05,nom06,tom06},
except for their large Co/Fe and Zn/Fe ratios (\eg
\citealt{dep02,bes05}). The enhancement of Co and Zn in low
metallicity stars requires explosive nucleosynthesis under high
entropy. In a ``spherical'' model, a high-entropy explosion corresponds to a
high-energy explosion that inevitably synthesizes a large amount of
\Nifs. One possible solution to this incompatibility is that some faint
SNe are associated with a narrow jet within which a high-entropy region
is confined \citep{ume05}. If this would be a realistic model for the
EMP stars, some faint SNe would accompany sufficiently energetic jets to
produce GRBs.

In this Letter, we present hydrodynamical and nucleosynthetic models of
the 40 $\Msun$ star explosions with relativistic jets.\footnote[10]{
Details of the numerical method, input physics, and the
performance of the code will be described in N. Tominaga (2007, PhD
thesis, in preparation).}  We show
that these models can explain the existence of GRB-HNe and GRBs without
bright SNe in a unified manner. We suggest that both GRB-HNe and GRBs
without bright SNe are BH-forming SNe involving relativistic jets and
that these explosions are responsible for the formation of the EMP stars.

\section{MODELS}
\label{sec:model}

We investigate the jet-induced explosions (\eg \citealt{mae03,nag06}) of
the $40\Msun$ Population III stars \citep{ume05,tom06} using a multi-dimensional
special relativistic Eulerian hydrodynamic code (\citealt{ume05b};
N.~Tominaga, 2007, PhD thesis, in preparation). 
We assume that the explosions of the $40\Msun$ Population III 
stars involve relativistic jets in analogy with the $40\Msun$ Pop I GRB-HNe, 
since the Fe core masses of Population III and Population I stars are similar. 

We inject the jets at a radius $R_{\rm in} \sim 900$ km, corresponding to an
enclosed mass of $M \sim 1.4 \Msun$, and follow the jet
propagation. Since the explosion mechanism of GRB-HNe is still under
debate, the jets are
treated parametrically with the following five parameters: energy
deposition rate ($\Ed$), total deposited energy ($E_{\rm dep}$), initial
half angle of the jets ($\theta_{\rm jet}$), initial Lorentz factor
(${\it \Gamma}_{\rm jet}$), and the ratio of thermal to total deposited
energies ($f_{\rm th}$). 

In this Letter, we investigate the dependence of nucleosynthesis outcome
on $\Ed$ for a range of $\Edep\equiv\Ed/10^{51}{\rm ergs\,s^{-1}}=0.3-1500$.
The diversity of $\Ed$ is consistent with the wide range of the observed
isotropic equivalent $\gamma$-ray energies and timescales of 
GRBs (\citealt{ama06} and references therein). 
Variations of activities of the central engines, possibly corresponding to
different rotational velocities or magnetic fields, may well produce 
the variation of $\Ed$. 
We expediently fix the other parameters as 
$E_{\rm dep}=1.5\times10^{52}$ergs, 
$\theta_{\rm jet}=15^\circ$, 
${\it \Gamma}_{\rm jet}=100$, and 
$f_{\rm th}=10^{-3}$ in all models. 

The thermodynamic history is traced with marker particles representing
individual Lagrangian elements (\eg \citealt{mae03}). The
nucleosynthesis calculation is performed as post-processing and includes
280 species up to $^{79}$Br (see \citealt{ume05}, Table~1). Because
recent studies (\eg \citealt{ram00,fro06}) suggest that
the electron fraction ($Y_{\rm e}$) is affected by neutrino
processes, we assume 
$Y_{\rm e}=0.5001$ in the Fe core to maximize the Co/Fe ratio \citep{ume05}.

The thermodynamic history of the jet is also traced by the marker
particles. In computing the jet composition, we assume that the jet
expands adiabatically from the Schwartzshild radius to the inner
boundary of the computational domain, and that the jet initially has the
composition of the accreted stellar materials. The details will be
presented in future work.

\section{RESULTS}
\label{sec:result}

In order to inject the jet, the ram pressure of the jet ($P_{\rm jet}$)
should overcome that of the infalling material ($P_{\rm fall}$). 
$P_{\rm jet}$ is determined by $R_{\rm in}$, $\Ed$, $\theta_{\rm jet}$, 
${\it \Gamma}_{\rm jet}$, and $f_{\rm th}$, thus being constant in time in
the present models. On the other hand, $P_{\rm fall}$ decreases with time,
since the density of the outer materials decreases following
the gravitational collapse (\eg \citealt{fry03}). For lower $\Ed$,
$P_{\rm jet}$ is lower, so that the jet injection 
($P_{\rm jet}>P_{\rm fall}$) is realized at a later time when the
central remnant becomes more massive due to more infall. 

\begin{figure}
\plotone{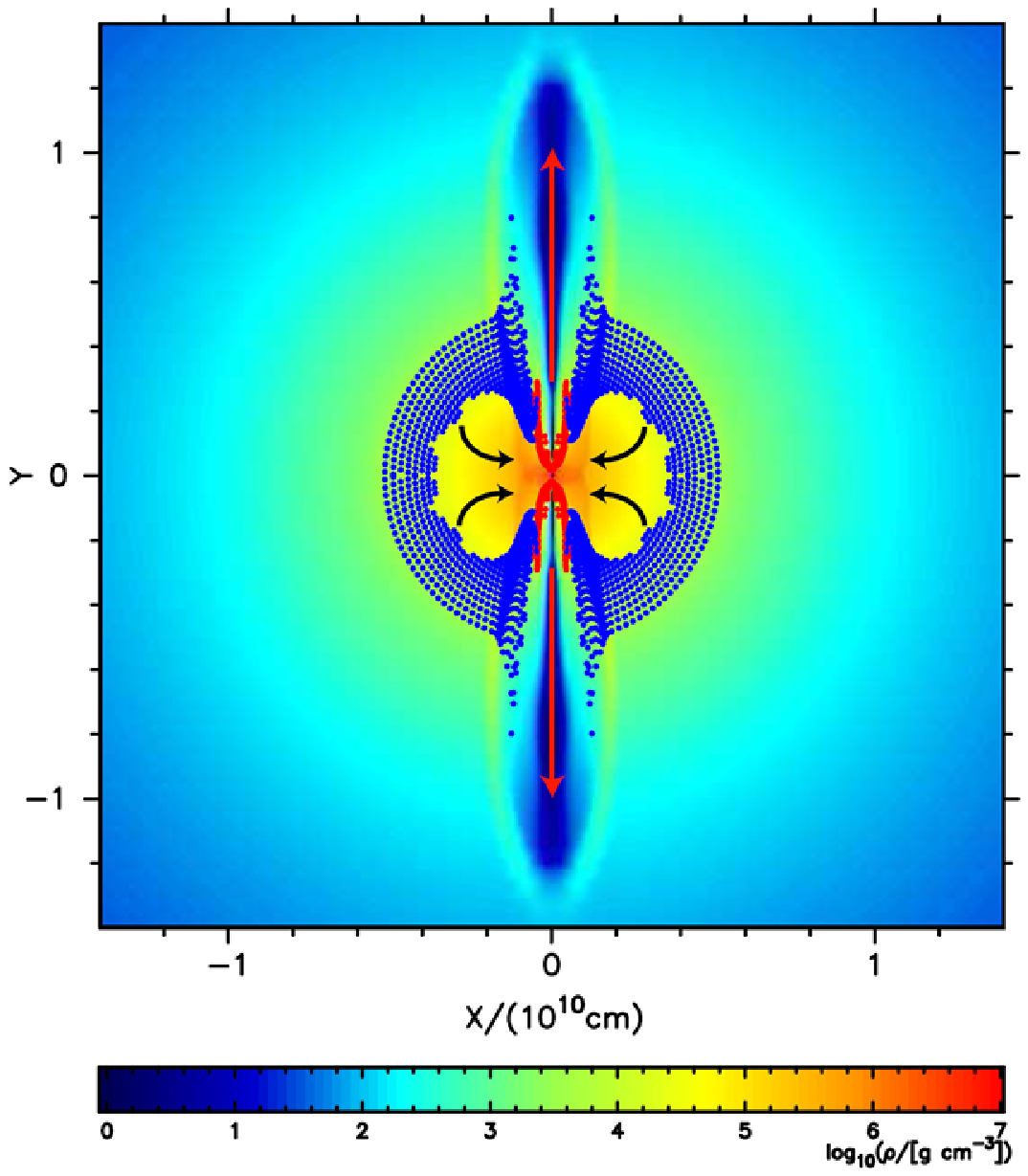}
\caption{Density structure of the 40 $\Msun$ Population III star
 explosion model of $\Edep=15$ at 1 s after the start of the jet
 injection. The jets penetrate the stellar mantle ({\it red arrows}) and
 material falls on the plane perpendicular to the jets ({\it black arrows}). 
 The dots represent ejected Lagrangian elements dominated by Fe 
 (\Nifs, {\it red}) and by O ({\it blue}). 
\label{fig:fallback}}
\end{figure}

After the jet injection is initiated, the shock fronts between the jets and
the infalling material proceed outward in the stellar
mantle. Figure~\ref{fig:fallback} is a snapshot of the model with
$\Edep=15$ at 1 s after the start of jet injection. When the jet
injection ends, the jets have been decelerated by collisions with the
dense stellar mantle and the shock has become more spherical. The inner
material is ejected along the jet-axis but not along the equatorial
plane. On the other hand, the outer material is ejected even along the equatorial
plane because the infall along the
equatorial plane is terminated as the shock reaches the equatorial
plane (\eg \citealt{mae03,nag06}). 

\subsection{\Nifs\ Mass}

\begin{figure}
\plotone{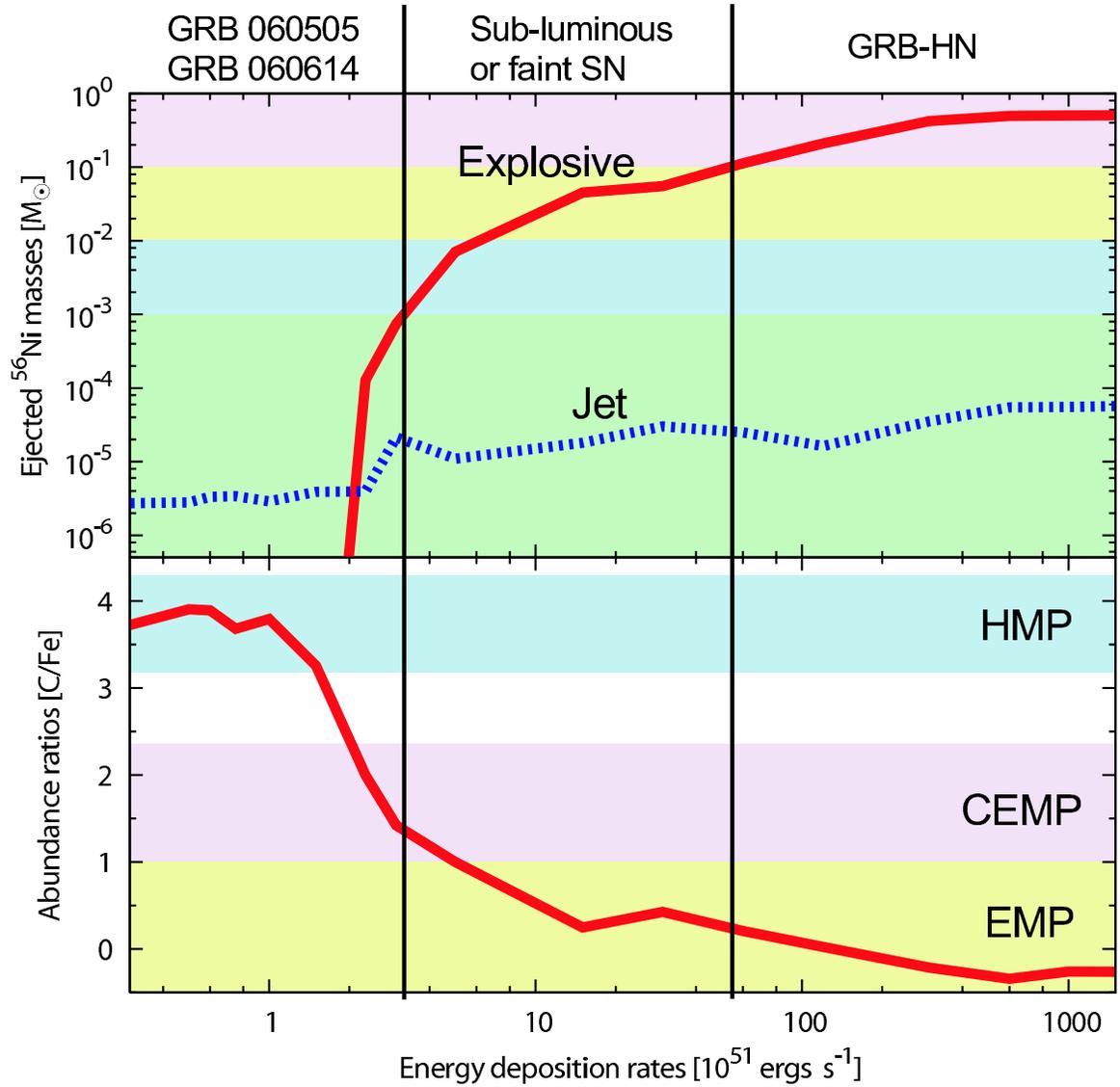}
\caption{{\it Top}: Ejected \Nifs\ mass ({\it red}: 
 explosive nucleosynthesis products, {\it blue}: the jet contribution)
 as a function of the energy deposition rate. The background color shows the
 corresponding SNe ({\it red}: GRB-HNe, {\it yellow}: sub-luminous SNe,
 {\it blue}: faint SNe, {\it green}: GRBs~060505 and 060614). 
 Vertical lines divide the resulting SNe according to their brightness. 
 {\it Bottom}: Dependence of abundance ratio [C/Fe] on the energy
 deposition rate. The background color shows the corresponding metal-poor 
 stars ({\it yellow}: EMP, {\it red}: CEMP, {\it blue}: HMP stars). 
\label{fig:EdotNi}}
\end{figure}

The top panel of Figure~\ref{fig:EdotNi} shows the dependence of the
ejected $\Mni$ on the energy deposition rate $\Ed$. For lower $\Ed$,
smaller $\Mni$ is synthesized in explosive nucleosynthesis because of
lower post-shock densities and temperatures (\eg
\citealt{mae03,nag06}; Maeda \& Tominaga 2007, hereafter \cite{mae06c}). 
While the materials in the C+O layer fall through the inner boundary,
$P_{\rm fall}$ decreases only moderately, being almost constant 
($P_{\rm fall}\sim P_{\rm fall}^{\rm C+O} \sim 10^{26}$ ${\rm dyn~cm^{-2}}$) 
during this phase because of the relatively shallow pre-SN density
gradient in the C+O layer (\eg \citealt{fry03,mae06c}). 

If $\Edep\gsim3$, $P_{\rm jet}>P_{\rm fall}^{\rm C+O}$ so that the jet
injection is initiated below the bottom of the C+O layer, leading to the
synthesis of $\Mni\gsim10^{-3}\Msun$. If $\Edep<3$, on the other hand,
$P_{\rm jet}<P_{\rm fall}^{\rm C+O}$ so that the jet injection is
delayed until $P_{\rm fall}$ decreases below $P_{\rm jet}$ and initiated
near the surface of the C+O core; then the ejected \Nifs\ is as small as
$\Mni<10^{-3}\Msun$.

\Nifs\ contained in the relativistic jets is only
$\Mni\sim10^{-6}$ to $10^{-4}\Msun$ because the total mass of the jets is
$M_{\rm jet}\sim 10^{-4}\Msun$ in our model with 
${\it \Gamma}_{\rm jet} = 100$ and $E_{\rm dep}=1.5\times10^{52}$ergs. 
The \Nifs\ production in the jets is predominant only for $\Edep<1.5$ in the
present model (Fig.~\ref{fig:EdotNi}). The models cannot synthesize
\Nifs\ explosively and eject very little $\Mni$ ($\sim10^{-6}\Msun$).

\subsubsection{GRB-HNe}

For high energy deposition rates ($\Edep\gsim60$), the explosions
synthesize large $\Mni$ ($\gsim0.1\Msun$) being consistent with GRB-HNe. 
The remnant mass was $M \sim1.4\Msun$ when the jet
injection was started, but it grows as material is accreted from the
equatorial plane. The final BH masses range from $M_{\rm BH}=10.8\Msun$ for
$\Edep=60$ to $M_{\rm BH}=5.5\Msun$ for 
$\Edep=1500$, which are consistent with the masses of
stellar-mass BHs \citep{bai98}. The model with $\Edep=300$
synthesizes $\Mni\sim0.4\Msun$ and results in $M_{\rm BH}=6.4\Msun$.  

Since the jet injection with large $\Ed$ is in short timescale and
terminated before the jets reach the surface of the C+O core, the
asphericity of the ejecta inside the C+O core for $\Edep\gsim60$ is
probably consistent with a relatively oval explosion as indicated by the
light curve and spectra of GRB-HNe \citep{mae06a,mae06b}. 

Neutrino annihilation has been estimated to provide only a small energy
deposition rate (\eg $\Edep\sim1$; \citealt{woo93}), leading to the
synthesis of smaller $\Mni$ than is required for GRB-HNe. In order to
synthesize a sufficient amount of \Nifs, large $\Ed$ should be produced via 
some mechanism, \eg magneto-rotation (\citealt{miz04}; another possibility, 
$^{56}$Ni production in the disk wind, is discussed in \citealt{mae06c}).

\subsubsection{GRBs~060505 and 060614}

For low energy deposition rates ($\Edep<3$), the ejected
\Nifs\ masses [$\Mni<10^{-3}\Msun$] are smaller than the upper
limits for GRBs~060505 and 060614. The final BH masses range from 
$M_{\rm BH}=18.2\Msun$ for $\Edep=0.3$ to 
$M_{\rm BH}=15.1\Msun$ for $\Edep=3$.
While the material ejected along the jet direction involves those from the
C+O core, the material along the equatorial plane is ejected only from the 
outer part of the H envelope. Thus $M_{\rm BH}$ exceeds the C+O core mass
$M_{\rm C+O}=13.9\Msun$ for the $40\Msun$ star.

If the star lost the H and He envelopes before its core collapsed,  
and if the explosion is viewed from the jet direction, we would observe 
GRB without SN re-brightening. This may be the situation 
for GRBs~060505 and 060614.

\subsubsection{GRBs with Faint or Subluminous SNe}

For intermediate energy deposition rates ($3\lsim\Edep<60$), the
explosions eject $10^{-3}\Msun \lsim \Mni <0.1\Msun$ and the final BH
masses are $10.8\Msun\lsim M_{\rm BH}< 15.1\Msun$. The resulting
SN is faint [$\Mni <0.01\Msun$] or sub-luminous 
[$0.01\Msun \lsim \Mni <0.1\Msun$]. 

Nearby GRBs with faint or sub-luminous SNe have not been observed.
Possible reasons may be that (1) they do not occur
intrinsically, \ie the energy deposition rate is bimodally distributed,
or that (2) the number of observed nearby GRBs is still too small. For
case 1, the GRB progenitors may be divided into two groups, \eg with
rapid or slow rotation and/or with strong or weak magnetic fields. For
case 2, future observations will detect GRBs
associated with a faint or sub-luminous SN.

\subsection{Abundance Ratio: {\rm C/Fe}}

The bottom panel of Figure~\ref{fig:EdotNi} shows the dependence of the
abundance ratio [C/Fe] on $\Ed$. Lower $\Ed$ yields larger $M_{\rm BH}$ and
thus larger [C/Fe], because the infall decreases the amount of inner core
material (Fe) relative to that of outer material (C) (see also
\citealt{mae03}). As in the case of $\Mni$ (\S 3.1), [C/Fe] changes
dramatically at $\Edep\sim3$. 

The abundance patterns of the EMP stars are good indicators of
nucleosynthesis in a single SN because the Galaxy was effectively unmixed at [Fe/H]
$<-3$ (\eg \citealt{tum06}). They are classified into three groups according to [C/Fe]: 
(1) [C/Fe] $\sim 0$, normal EMP stars ($-4<$ [Fe/H] $<-3$, \eg \citealt{cay04}); 
(2) [C/Fe] $\gsim+1$, Carbon-enhanced EMP (CEMP) stars ($-4<$ [Fe/H] $<-3$, 
\eg CS~22949-37; \citealt{dep02}); 
(3) [C/Fe] $\sim +4$, hyper metal-poor (HMP) stars ([Fe/H] $<-5$, 
\eg HE~0107--5240; \citealt{chr02}; \citealt{bes05}; HE~1327--2326, \citealt{fre05}).

Figure~3 shows that the general abundance patterns of the normal EMP
stars, the CEMP star CS~22949-37, and the HMP stars HE~0107--5240 and
HE~1327--2326 are reproduced by models with $\Edep=120$, 3.0, 1.5, and
0.5, respectively (see Table~1 for model parameters). The model for the
normal EMP stars ejects $\Mni\sim0.2\Msun$, i.e. a factor of 2 less than
SN~1998bw. On the other hand, the models for the CEMP and the HMP stars
eject $\Mni\sim8\times10^{-4}$ and $4\times 10^{-6}\Msun$,
respectively, which are always smaller than the upper limits for
GRBs~060505 and 060614. The lack of the metal-poor stars at $-5<$ [Fe/H]
$<-4$ is explained by the narrow range of $\dot{E}_{\rm dep}$. The N/C
ratio in the models for CS~22949-37 and HE~1327--2326 is enhanced by
partial mixing between the He and H layers during pre-SN evolution (\citealt{iwa05}).

\section{DISCUSSION AND CONCLUSION}
\label{sec:discuss}

We have computed hydrodynamics and nucleosynthesis for the explosions induced
by relativistic jets. We have shown that (1) the
explosions with large $\Ed$ are observed as GRB-HNe and their yields explain the
abundances of normal EMP stars, and (2) the explosions with small $\Ed$
are observed as GRBs without bright SNe and are responsible for the
formation of the CEMP and the HMP stars. We thus propose that GRB-HNe
and GRBs without bright SNe belong to a continuous series of BH-forming
SNe with the relativistic jets of different $\Ed$.

Presently, the number fraction of GRBs without bright SNe relative to
the known nearby GRBs is $\sim40$\% (\eg \citealt{woo06}). 
On the other hand, among the EMP stars with [Fe/H] $<-3.5$, the
fractions of the CEMP and the HMP stars relative to the EMP stars are
$\sim25\%$ and $\sim15\%$, respectively (\citealt{bee05}). Although the
numbers of observed GRBs and the EMP stars with [Fe/H] $<-3.5$ are still
too small to discuss statistics, the fraction of GRBs without bright SNe is
consistent with the sum of the fractions of the CEMP and the HMP
stars. Thus GRBs~060505 and 060614 are likely related to the CEMP stars
ejecting $\Mni\sim 10^{-4}$ to $10^{-3}\Msun$ or HMP stars ejecting 
$\Mni\sim 10^{-6}\Msun$. 

A short GRB, probably the result of the merger of two compact objects 
(\eg \citealt{geh05}), synthesizes virtually no \Nifs\
because the ejecta must be too neutron-rich. 
In contrast, our model suggests that GRBs~060505 and 060614 
produced $\Mni\sim10^{-4}-10^{-3}\Msun$ or $\sim 10^{-6}\Msun$. 
If such a GRB without a bright SN occurs in a very faint and nearby
galaxy, our model predicts that some re-brightening due to the \Nifs\
decay can be observed.

\begin{deluxetable}{ccccc}
 \tablecaption{Models compared with metal-poor stars.\label{tab:model}}
 \tablewidth{0pt}
 \tablehead{
   \colhead{Stars}
 & \colhead{$\Ed$}
 & \colhead{$\Mni$}
 & \colhead{$M_{\rm BH}$}
 & \colhead{[C/Fe]} \\
   \colhead{}
 & \colhead{[$10^{51} {\rm ergs~s^{-1}}$]}
 & \colhead{[$\Msun$]}
 & \colhead{[$\Msun$]}
 & \colhead{}
 }
\startdata
EMP           &  120  & $2.09\times10^{-1}$ &  $9.1$ & 0.02 \\
CS~22949--37  &  3.0  & $7.64\times10^{-4}$ & $15.1$ & 1.20 \\
HE~1327--2326 &  1.5  & $3.90\times10^{-6}$ & $16.9$ & 3.21 \\
HE~0107--5240 &  0.5  & $2.80\times10^{-6}$ & $17.1$ & 3.91 \\
\enddata
\end{deluxetable}

\clearpage

\clearpage

\clearpage

\begin{figure}
\plotone{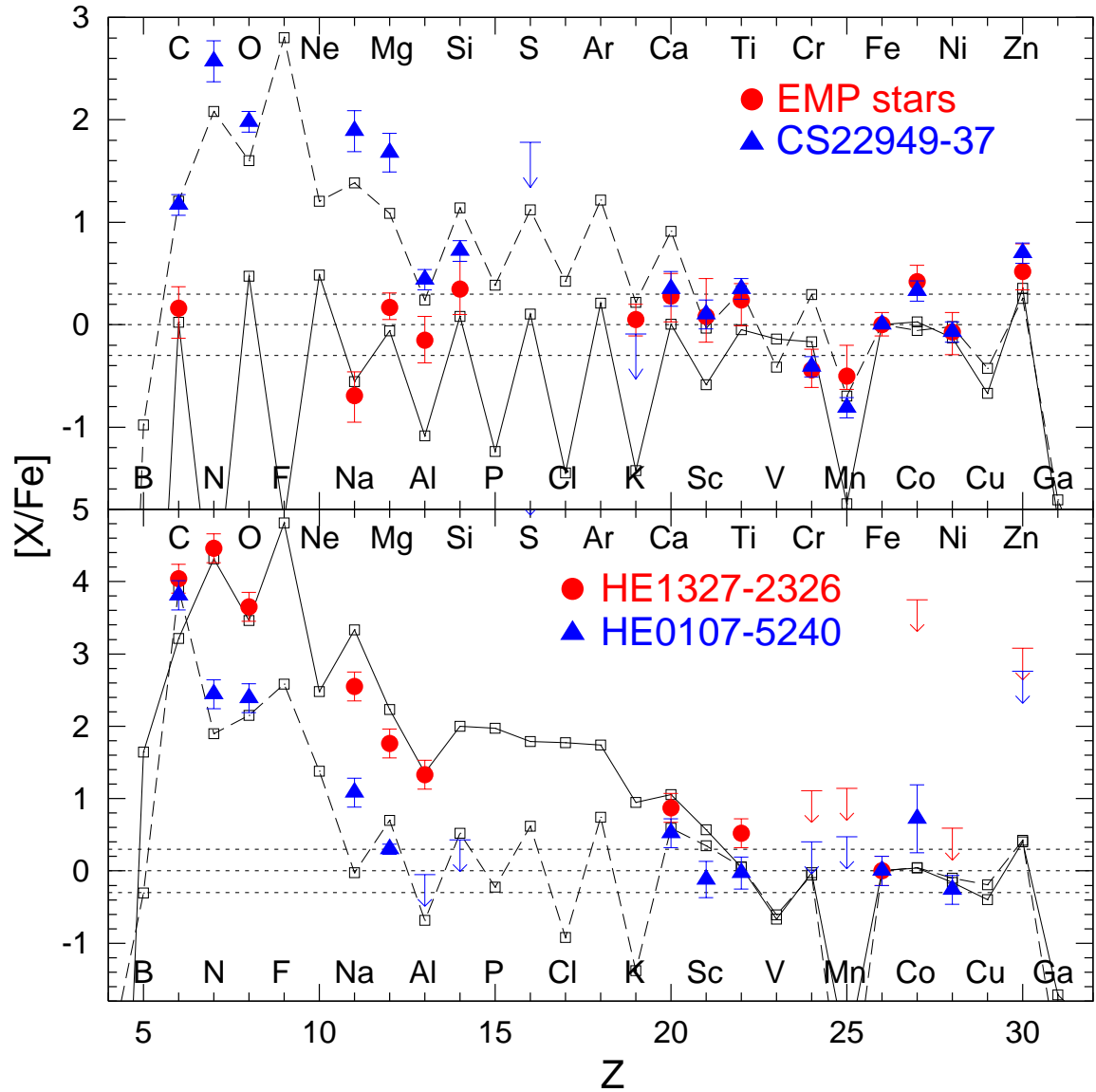}
\caption{Comparison of the abundance patterns of metal-poor 
 stars and of models. 
 {\it Top}: Normal EMP ({\it red dots}) and 
 CEMP ({\it blue triangles}) stars and models
 with $\Edep=120$ ({\it solid line}) and $=3.0$ ({\it dashed line}).
 {\it Bottom}: HMP stars:
 HE~1327--2326, ({\it red dots}), 
 and HE~0107--5240, ({\it blue triangles}) and models 
 with $\Edep=1.5$ ({\it solid line}) and $=0.5$ ({\it dashed line}).
\label{fig:EMP}}
\end{figure}

\end{document}